\documentclass[9pt,twocolumn,twoside]{article}
\usepackage{amsmath}
\usepackage{authblk}
\usepackage{graphicx}% Include figure files
\usepackage{dcolumn}% Align table columns on decimal point
\usepackage{bm}% bold math
\usepackage{cite}
\usepackage[margin=.6in]{geometry}

\title{Ultrafast tunable modulation of light polarization at terahertz frequencies}

\author[1,*]{Vincent Juv\'e}
\author[1]{Gwena\"elle Vaudel}
\author[2,3]{Zoltan Ollmann}
\author[2]{Janos Hebling}
\author[1]{Vasily Temnov}
\author[4]{Vitalyi Gusev}
\author[1]{Thomas Pezeril}

\affil[1]{Institut des Mol\'ecules et Mat\'eriaux du Mans, UMR CNRS 6283, 
Le Mans Universit\'e, avenue Olivier Messiaen, 72085 Le Mans, France}
\affil[2]{Department of Experimental Physics and MTA-PTE High Field Terahertz Research Group, University of P\'ecs, 7624 P\'ecs, Hungary}
\affil[3]{Presently at Institute of Applied Physics, University of Bern, Bern, CH 3012, Switzerland}
\affil[4]{Laboratoire d'Acoustique de l'Universit\'e du Maine, UMR CNRS 6613,
Le Mans Universit\'e, avenue Olivier Messiaen, 72085 Le Mans, France}
\affil[*]{Corresponding author: vincent.juve@univ-lemans.fr}

\begin{document}

\maketitle

\textbf{Controlling light polarization is one of the most essential
routines in modern optical technology. Since the demonstration
of optical pulse shaping by spatial light modulators and
its potential in controlling the quantum reaction pathways,
it paved the way for many applications as a coherent control
of the photoionization process or as polarization shaping of
terahertz (THz) pulses. Here, we evidenced efficient nonresonant
and noncollinear $\chi^{(2)}$-type light-matter interaction
in femtosecond polarization-sensitive time-resolved optical
measurements. Such nonlinear optical interaction of visible
light and ultrashort THz pulses leads to THz modulation of
visible light polarization in bulk LiNbO$_3$ crystal. Theoretical
simulations based on the wave propagation equation capture
the physical processes underlying this nonlinear effect.
Apart from the observed tunable polarization modulation
of visible pulses at ultrahigh frequencies, this physical
phenomenon can be envisaged in THz depth-profiling of
materials.}

Over the last decade, ultrafast terahertz (THz) spectroscopy has
gained tremendous attention thanks to the development of high power
ultrafast laser systems~\cite{kampfrath2013resonant}, which allowed the generation of
intense single-cycle picosecond pulses of electric field at THz
frequencies~\cite{yeh2007generation,hirori2011single}. Their relatively long optical cycle period
(1 ps for 1 THz) and high electric field (from hundreds of
kV/cm to a few MV/cm) provide a new tool for studying fundamental
aspects of light-matter interactions like THz-driven
ballistic charge currents in semiconductors~\cite{kuehn2010coherent,schubert2014sub}, linear~\cite{kampfrath2011coherent} and nonlinear~\cite{baierl2016nonlinear} control of antiferromagnetic spin waves, THz-induced dynamics in metallic ferromagnets~\cite{vicario2013off} and transparent magneto-optical media~\cite{subkhangulov2016terahertz}, or even an ultrafast switching of electric polarization in ferroelectrics~\cite{mankowsky2017ultrafast}. Field-resolved detection
of ultrashort THz pulses has been well known for many
years, and the most common technique is based on free-space
electro-optic sampling, which can be implemented in two ways:
non phase-matched electro-optic sampling in crystals like ZnTe~\cite{planken2001measurement} or GaP~\cite{wu19977} and phase-matched electro-optic sampling in a GaSe crystal~\cite{kubler2004ultrabroadband,liu2004gase}. The latter relies on phase-matched propagation
of THz and visible pulses in a crystal leading to an efficient
ordinary-to-extraordinary conversion for visible photons at approximately
the same frequency, $\hbar\omega_{\rm e}=\hbar\omega_{\rm o}\pm\hbar\omega_{\rm o}^{\rm THz}\simeq\hbar\omega_{\rm o}$. This leads to the polarization change of an optical pulse, which is
detected by the polarization-sensitive scheme, and to the field resolved
detection of a THz pulse. Since nonresonant THz induced
optical nonlinearities in dielectric crystals are nearly
instantaneous, the magnitude of the polarization change should
be determined solely by the spatiotemporal dynamics of interacting
THz and visible pulses propagating inside the crystal.
Consequently, the polarization switching time lies in the picosecond
range governed by the duration of the pulses involved. In
other words, such nonresonant nonlinear optical interactions set
the fundamental speed limits of electro-optical modulation,
which is limited to nanosecond switching time for conventional
electro-optic devices~\cite{lu2001wide}. Although phase-matched electro-optics
seems to be appealing to achieve ultrafast control over
the polarization state of light, to the best of our knowledge, it
has been only applied to detect and characterize THz fields.
Moreover, the relatively low efficiency and the third-order optical
nonlinearity in a GaSe crystal are limiting its impact~\cite{sell2008phase}.
\begin{figure}[!ht]
\centering
\includegraphics[trim=0cm 2.5cm 0cm 0.1cm, clip=true,width=\columnwidth]{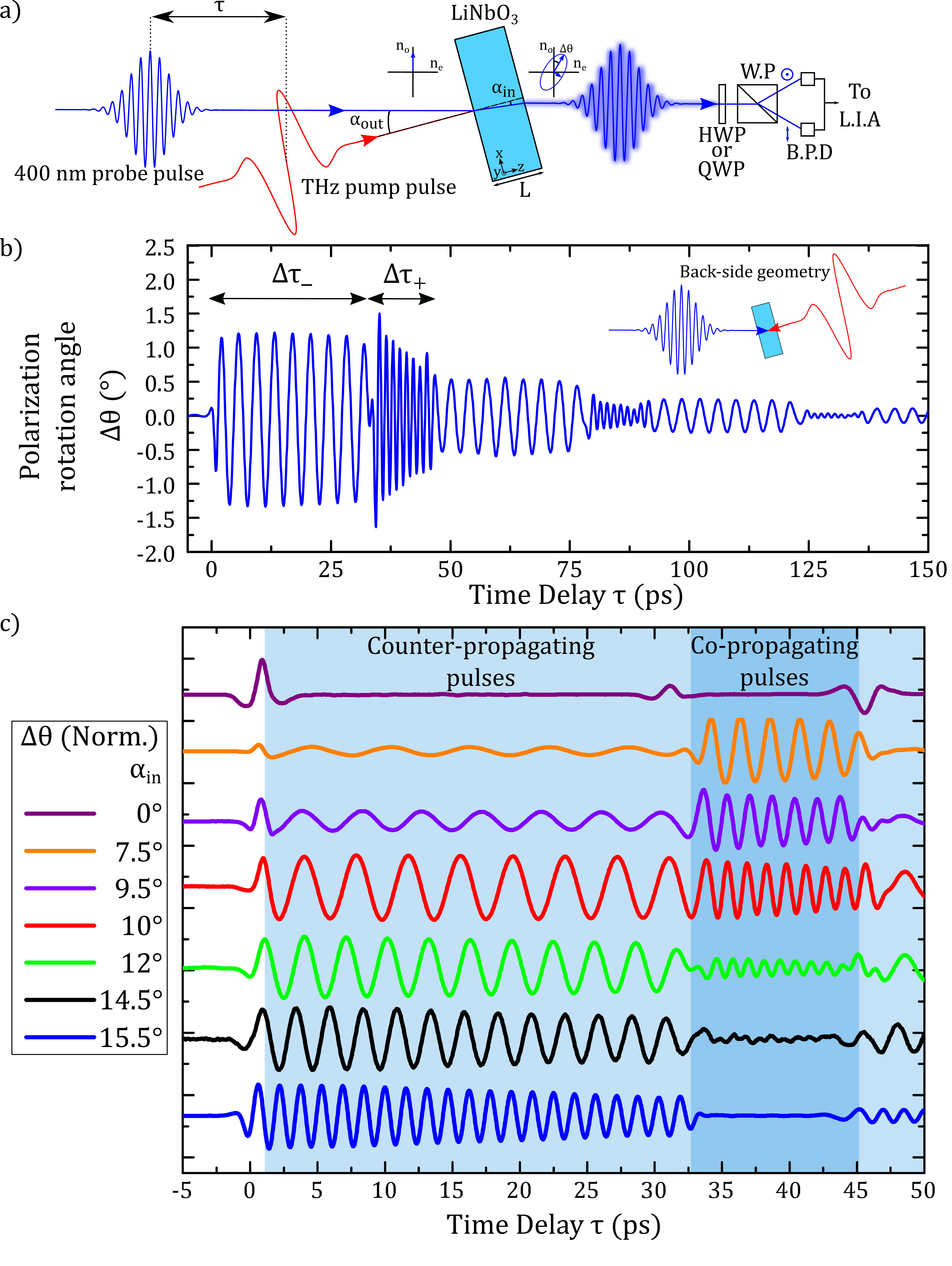}%the trim option is necessary for the new version of this fig.
\caption{(a) Schematic of the time-resolved polarization change
measurement in a z-cut LiNbO$_3$ crystal in the front-side probing
configuration. HWP, half-wave plate; QWP, quarter-wave plate. WP,
Wollaston prism. BPD, balanced photodiodes. LIA, lock-in amplifier.
L, sample’s thickness. (b) Relative change of the probe light polarization
orientation as a function of the time delay τ between the THz pump
pulse and the 400 nm probe pulse for $\alpha_{\rm out}=25^\circ$($\alpha_{\rm in}=10^\circ$) in the
back-side probing configuration, shown by the small sketch.
(c) Ultrafast transients obtained for different probe angles as a function
of the time delay between the pump and the probe in the back-side
probing configuration. The curves are normalized and shifted for
the sake of clarity.
\label{fig:setup}}
\end{figure}

Here we report on the efficient modulation of light's polarization
by THz ultrashort pulses via the second-order nonlinear
optical interaction inside a LiNbO$_3$ crystal, where an intense
THz pump and a visible probe pulse interact during their
propagation inside the crystal of thickness 1 mm. Optical properties
of visible probe pulses are monitored by polarization sensitive
measurements (rotation and ellipticity) as a function
of the pump-probe delay time. Two distinct (front-side and
back-side) probing configurations are sketched in Fig.~\ref{fig:setup}~(a) and in the inset of Fig.~\ref{fig:setup}~(b), respectively. The maximum THz electric field amplitude on the sample is in the order of 100 kV/cm. Fig.~\ref{fig:setup}~(b) shows a typical pump-probe transient measured in the back-side probing configuration at 400~nm probe wavelength and $\alpha_{\rm in}=10^\circ$ ($\alpha_{\rm out}=25^\circ$). Due to the polarization-sensitive detection scheme with a half waveplate, the measured signal is linked to the rotation of the polarization state of the visible probe pulses induced by nonlinear optical interaction with the THz pulses propagating in the crystal and exhibits time domain oscillations at two distinct frequencies of 0.22~THz (low frequency $\nu_-$) and 0.69~THz (high-frequency $\nu_+$), respectively. For this particular experimental configuration, the rotation angle, denoted as the polarization rotation angle $\Delta\theta$, is approximately $1^\circ$, which corresponds to a polarization conversion efficiency of $\sim 5\%$. A similar
transient shape is observed when measuring the ellipticity
change with a quarter-wave plate (not shown). Moreover, the
signals are of comparable amplitude when measuring the rotation
or ellipticity changes, but they are phase-shifted by roughly
$\pi/2$, which means that the rotation is maximum when the
ellipticity is 0 and vice versa. The occurrence of high-frequency
$\nu_+$ and low-frequency $\nu_-$ can be uniquely attributed to copropagating
and counterpropagating THz and visible pulses,
respectively. The frequency changes from high to low, and vice
versa, each time the THz pulse is reflected at one of the crystal
surfaces. The alternating time windows for the observation of copropagating and counterpropagating THz and visible pulses
inside the crystal of thickness L= 1 mm (see Fig~\ref{fig:setup}~(b)) are
governed by group velocities, 
\begin{equation}
\Delta\tau_+\approx\frac{L}{c}(n_{\rm THz}- n_{\rm o}^{\rm gr}) {\ \rm and\ } \Delta\tau_-\approx\frac{L}{c}(n_{\rm THz}+ n_{\rm o}^{\rm gr})\, ,
\label{eq:tau}
\end{equation}
with $n_{\rm o}^{\rm gr}\sim 3.06$ the index of light polarized along the ordinary and $n_{\rm THz}^{\rm gr}=n_{\rm THz}$ the refractive index of the THz pulse, considering that the LiNbO$_3$ is not dispersive in this frequency range. The different time intervals $\Delta\tau_{\rm +,-}$ are experimentally found to be 12.1~ps and 32.3~ps, and, considering the value of the visible group refractive index, one can extract the THz ordinary refractive index $n_\textrm{THz}\approx6.75$, which is in good agreement with the literature~\cite{schall1999far,unferdorben2015measurement}. Finally the signal duration of hundreds of picoseconds and the drops in amplitude for each reflection at the interfaces are explained by the absorption coefficient of the LiNbO$_3$ in the THz spectral range and by the THz reflection coefficient, which is roughly 75\% for the field amplitude.

\begin{figure}[!ht]
\centering
\includegraphics[trim=.0cm .0cm .0cm .0cm, clip=true,width=\columnwidth]{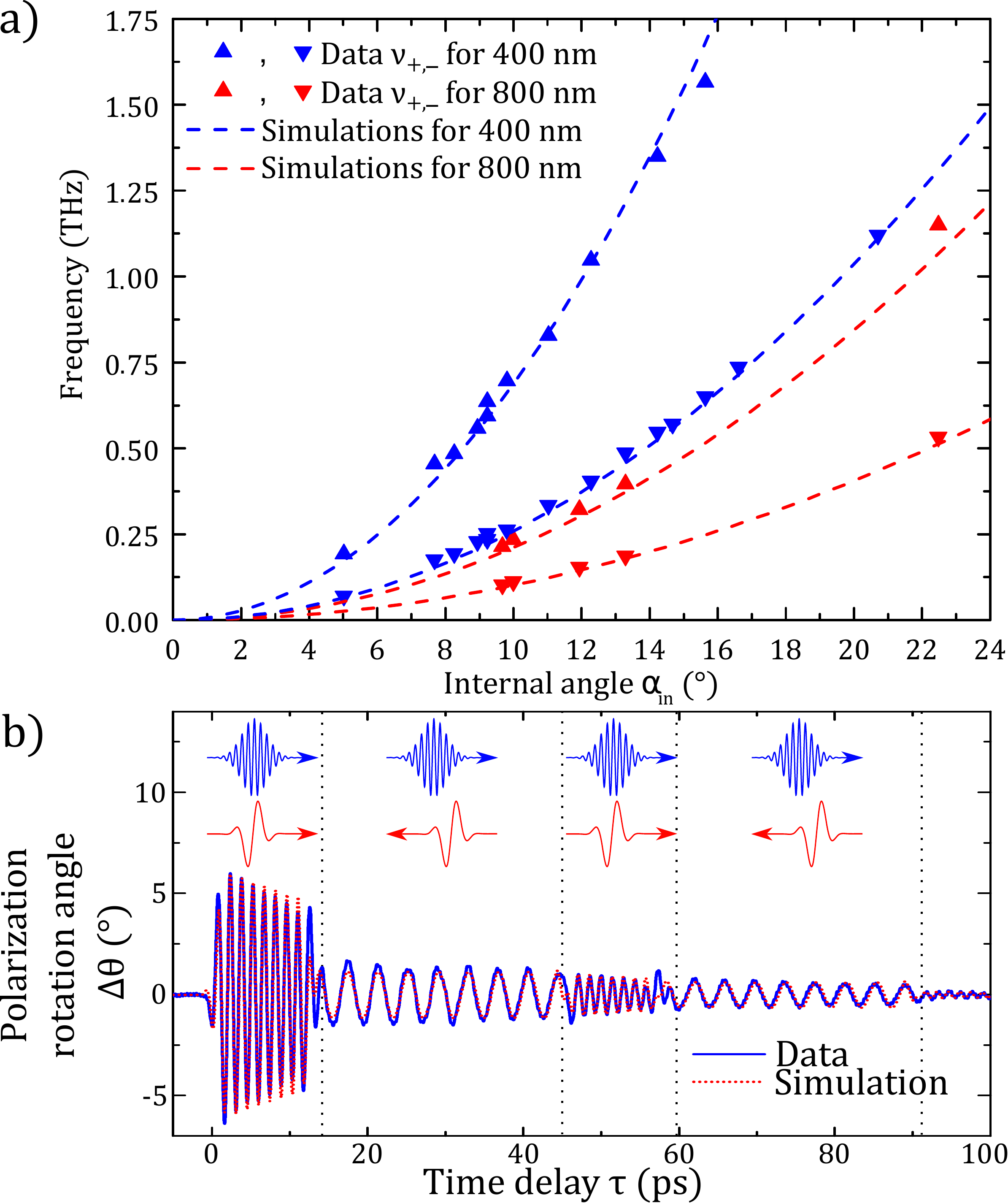}
\caption{(a) Measured frequencies as a function of the probe’s incidence
angle. Two different probing wavelengths (triangles) are plotted
as well as the predictions of our model for both wavelengths (dashed
lines). (b) Comparison of the experimental and simulated polarization rotation transients for an angle $\alpha_{\rm in}=10^\circ$ and $\lambda_{\rm o}=400$~nm. The corresponding oscillations frequencies are $\nu_+=0.69$~THz and $\nu_-=0.27$~THz. Vertical dotted lines separate the regions with co- and counter-propagating pump and probe pulses.  \label{fig:FreqVSAngle}} 
\end{figure}

Measurements at different probe angles $\alpha_{\rm in}$ ranging from 0 to 16$^\circ$ have been performed and are displayed in Fig.~\ref{fig:setup}~(c) for a probe's wavelength $\lambda_{\rm o}=400$~nm. The curve at 0$^\circ$ reflects interfacial effect as described below. On top of this effect, noncollinear geometry gives birth to the single low and high frequencies, which increase continuously as a function of the angle $\alpha_{\rm in}$. Systematic investigation of the evolution of the observed frequencies versus the probe angles $\alpha_{\rm in}$ and for two wavelengths $\lambda_{\rm o}=400$~nm and $\lambda_{\rm o}=800$~nm was carried out and is displayed in Fig.~\ref{fig:FreqVSAngle}~(a). For each wavelength, two frequency branches ($\nu_+$ and $\nu_-$) appear, which correspond to  co- and counter-propagating THz and visible pulses as described previously. The maximum measured frequency is 1.6~THz ($\alpha_{\rm in}\approx15^\circ$) for $\lambda_{\rm o}=400$~nm and 1.1~THz ($\alpha_{\rm in}\approx22^\circ$) for $\lambda_{\rm o}=800$~nm. Dashed lines correspond to the results of the theoretical solution based on our model, which is summarized by Eq.~(\ref{eq:wavetheorysolution_intensityCoPro}), and are perfectly in agreement with experimental results. Fig.~\ref{fig:FreqVSAngle}~(b) demonstrates perfect agreement between the experimental and simulated polarization rotation in time domain.

\begin{figure}[!b]
\centering
\includegraphics[trim=.0cm .0cm .0cm .2cm, clip=true,width=\columnwidth]{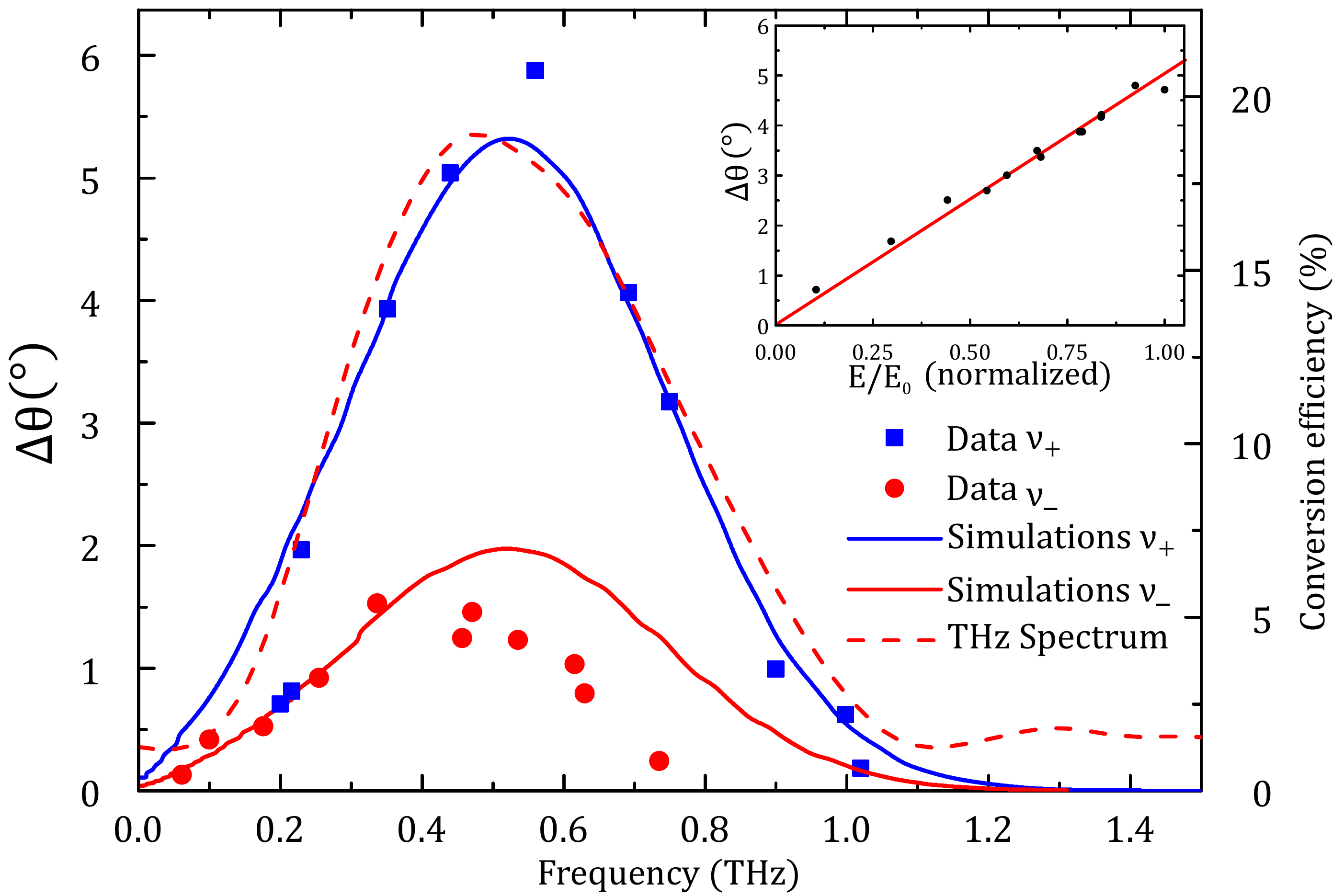}
\caption{Polarization rotation angle as a function of the measured
frequency for the high- and low-frequency modes. For comparison,
the experimental THz spectrum is shown in the dashed line. Inset,
polarization rotation angle for the high-frequency oscillation versus
the THz electric field strength for $\alpha_{\rm in}=9.3^\circ$ and $\lambda_{\rm o}=400$~nm in the front-side configuration. \label{fig:AmplitudeVSFreq}} 
\end{figure}

For a  quantitative insight into the nonlinear process involved here, we have measured the amplitude of the change of the light's polarization angle $\Delta\theta$ while sweeping the internal angle $\alpha_{\rm in}$ for a wavelength of 400~nm. In Fig.~\ref{fig:AmplitudeVSFreq}, $\Delta\theta$ of the first occurring oscillations ($\nu_+$ or $\nu_-$)  is plotted as a function of the frequencies for the two experimental configurations namely the front-side and back-side probing geometries. The first striking feature is the measured maximum polarization rotation angle, occurring at 0.6~THz for $\nu_+$, which is as high as $6^\circ$. This corresponds to a light polarization conversion efficiency of 20\% upon the presence of the THz pulse. This light's polarization rotation angle is the largest, for the considered
THz field amplitude, reported so far in the literature
(in the order of 1 degree in ultrafast experiments~\cite{vicario2013off,mikhaylovskiy2016colossal}). Furthermore, unlike GaSe~\cite{sell2008phase}, no saturation effect is evidenced (see the inset of Fig.~\ref{fig:AmplitudeVSFreq}), opening the way to improving conversion
efficiency. The second striking feature is the difference in the maximum rotation angle for $\nu_+$ and $\nu_-$ with a almost 3 times higher amplitude for $\nu_+$. Along with the experimental data points, the results of numerical evaluation of the analytical theoretical prediction of time integration of Eq.~(\ref{eq:wavetheorysolution_intensityCoPro}) are plotted for both co- ($\nu_+$) and counter-propagating ($\nu_-$) pulses. These are in a good agreement with the experimental results.
Moreover, the spectral amplitude of the polarization rotation
transients is proportional to the spectral components of the
THz pulse used in the experiments shown in the dashed line
in Fig.~\ref{fig:AmplitudeVSFreq}. In light of these results, the main effect of tuning
the angle is to fulfill the phase-matching condition for different
frequency components contained in the pump THz spectrum,
which directly leads to the THz pulse sampling in the frequency
domain and, thus, allows to adjust the modulation frequency of
the probe pulse.

The experimental results were analysed by theoretical
simulations of the scattering process between the light and
THz pulses within the sample. We considered second-order
nonlinear interaction between the electric field of the ordinary
polarized THz pulse and of the ordinary polarized visible probe,
leading to the generation of an extraordinary optical beam. Solving the wave propagation equation for the extraordinary
beam generated by this interaction leads to the expression for
the signal amplitude as the function of the time delay $\tau$ between
the THz pump and the optical probe beam, which is governed
by an interplay between distinct phase- and group-matching
conditions for visible and THz electromagnetic fields. For the
half-wave plate and Wollaston prism arrangement, the theoretical
modelling gives

\begin{equation}
\begin{aligned}
I^\pm_{\lambda/2}&(t,\tau,\alpha_{\rm in})\sim\frac{\pi {\rm d}_{22}n_{\rm e}n^2_{\rm o}}{2\lambda_{\rm o}}\times \\
&A_{\rm o}(t-\tau-Ln_{\rm o}^{\rm gr}/c)\int_0^L dzE_{\rm THz}\Big(t-(n_{\rm THz}\mp n_{\rm e}^{\rm gr})\frac{z}{c}\Big )\times \\
&A_{\rm o}\Big(t-\tau-(n_{\rm o}^{\rm gr}-n_{\rm e}^{\rm gr}) \frac{z}{c}\Big) \sin{ \Big[\frac{2\pi}{\lambda_{\rm o}}(n_{\rm e}-n_{\rm o})z+\Phi(\alpha_{\rm in})\Big] }\,,
\label{eq:wavetheorysolution_intensityCoPro}
\end{aligned}
\end{equation}

where $d_{22}$ is the nonlinear optical coefficient, $\Phi(\alpha_{\rm in})=(k_{\rm o}(\omega_{\rm o})-k_{\rm e}(\omega_{\rm o},\alpha_{\rm in}))L$ with $k_{\rm o,e}(\omega_{\rm o,e})=(2\pi/\lambda_{\rm o})n_{\rm o,e}$ is the wave vector for the ordinary or extraordinary polarized visible light respectively and $A_{\rm o}(t)$ is a slowly varying envelope of the optical probe pulse. $I^+(t,\tau,\alpha_{\rm in})$ and $I^-(t,\tau,\alpha_{\rm in})$ stand for co-propagating and counter-propagating pulses, respectively. The photodetector signal $S(\tau,\alpha_{\rm in})\sim\int _{-\infty}^{+\infty}I^\pm_{\lambda/2}(t,\tau,\alpha_{\rm in}) dt$ is the time integral of Eq.~(\ref{eq:wavetheorysolution_intensityCoPro}) and leads to the time evolution of the polarization state (rotation and ellipticity) of the probe beam. The simulated transient (see figure~\ref{fig:FreqVSAngle} (b)) gives information about the modulation frequency and the amplitude as a function of the angle $\alpha_{\rm in}$. Multiple reflections can also be simulated.
%(see Supplementary Material and Eqs.~S16 and~S17). 
In our
model, we only consider the second-order nonlinear effect occurring
in the bulk of the crystal. In the simulations, there are no
free parameters: the model requires only the parameters of the
material and of the pulse, which are determined independently.
The simulated transients are in excellent agreement with the experimental
ones as we included the reflection coefficient as well
as the THz absorption of the crystal (see Fig.~\ref{fig:FreqVSAngle}~(b)). The computed frequencies of the oscillations versus the angle $\alpha_{\rm in}$ are displayed in dashed lines in Fig.~\ref{fig:FreqVSAngle} and perfectly agree the experimental observation for two different wavelengths. There is as well a good agreement between the calculated and measured amplitudes of the oscillations, which are displayed in Fig.~\ref{fig:AmplitudeVSFreq} with only a scaling factor. It is interesting to note that the ratio of the $\nu_+$ and $\nu_-$ simulated amplitudes is frequency independent and is equal to 2.7. This ratio can be explained by the longer interaction time of the THz pulse with the co-propagating light field than with the counter-propagating one, %$\bigg(\frac{\Delta n^{\rm gr}_ -}{\Delta n^{\rm gr}_ +}\bigg)\approx 2.7$
$\bigg(\frac{n_{\rm THz}+ n_{\rm o}^{\rm gr}}{n_{\rm THz}- n_{\rm o}^{\rm gr}}\bigg)= 2.7$, agreeing with the experimental value of $\approx3$. Deviation of the numerical simulation starting at around 0.4~THz for the back-side configuration, corresponding to an angle $\alpha_{\rm in}\approx 15^\circ$ ($\alpha_{\rm out}\approx 38^\circ$), can be explained by the geometrical effect of the beams crossing in the sample which is neglected in our model.

We now discuss the physical effect underlying the THz controlled
polarization rotation of the visible probe pulse. In
conventional non phase-matched free-space electro-optic sampling,
in ZnTe crystal for instance, the THz and optical gate
pulses are group-matched. Thus, the optical probe pulse accumulates
a phase shift at a constant rate during the propagation
in the crystal proportional to the THz electric field value set by
the relative time delay between the pulses. This technique has been widely used for decades in linear and nonlinear THz spectroscopy~\cite{grischkowsky1990far} and provides a sampling of the THz pulse in time domain by the light pulse. Moreover, the effect exists both for collinear~\cite{van1989high,wu1995free} and noncollinear~\cite{shan2000single} propagation of THz and light pulses. In the case of LiNbO$_3$, the optical and THz group indices are mismatched by roughly a factor of 2, which leads to a different interpretation of our experimental results. The observed experimental signals have two origins: from the interface region and from the bulk. The effect of the interface region can be clearly seen in Fig.~\ref{fig:setup}~(c) for $\alpha_{\rm in}=0^\circ$ at $\tau=0,~32$ and 47~ps corresponding to the injection or reflection of THz pulses from the surface as seen by the optical probe. Consequently, the non-zero polarization rotation will be proportional to $\int_{-\infty}^{\tau} E_{\rm THz}(t) dt\neq 0$. Once the THz pulse is completely inside the crystal, the polarization rotation accumulated by the optical pulse will be proportional to the integral of the THz pulse $\int_{-\infty}^{+\infty} E_{\rm THz}(t) dt=0$ and, thus, leads to no detectable signal. It has to be noted that the observed "interfacial" effect exploits the substantial group mismatch between the optical and THz pulses (common for most materials) and can be used to retrieve the temporal THz pulse shape. The periodic oscillation of the signal relies on the complex interplay between phase and group-matching conditions (see Eq.~(\ref{eq:wavetheorysolution_intensityCoPro})) due to optical second order nonlinear effect occurring inside the sample and is observed only for noncollinear propagation of the THz and light pulses. Its frequency can be adjusted by tuning the angle between the two incident beams. The angular dependence of the oscillation amplitude makes possible to retrieve the THz spectrum as shown in Fig.~\ref{fig:AmplitudeVSFreq}. 

%\begin{figure}[ht]
%\centering
%\includegraphics[trim=0cm 0cm 0cm 0cm, clip=true,width=\columnwidth]{fig4}
%\caption{Three dimensional representation of the envelope of an optical pulse polarization shaped by nonlinear interaction with a THz pulse in a LiNbO$_3$ crystal for an input ordinary polarized optical pulse defined by its envelope $A_{\rm o}(t)\sim {\rm exp}(-4\text{ln}(2)t^2/\Delta t^2)$ with a pulse duration of $\Delta t=15$~ps. The polarization-shaped pulse envelope is defined by $\vec{A(t)}=A_{\rm o}(t)\vec{{\rm e_x}}+A_{\rm e}(t)\vec{{\rm e_y}}$ (see the Supplementary Materials). The polarization modulation frequency is 0.5 THz, the maximum polarization rotation is $10^\circ$ and the maximum ellipticity is 0.5. The value of the ellipticity change was blown up for sake of clarity. a) to d) represent the light's polarization state for different times t. Inset: Experimental transient of the rotation and ellipticity relative changes of the probe pulse for an angle $\alpha_{\rm in}\sim 9^\circ$ and $\lambda=400$~nm.}
%\label{fig:comparison}
%\end{figure}

One potential application would be to use picosecond pulses of visible light with a pulse duration roughly equal to $\Delta\tau_-$ or $\Delta\tau_+$. This configuration would lead to an ultrafast periodic modulation of the polarization state of the visible light at THz frequency, depending on the crossing angle $\alpha_{\rm in}$ between THz and optical pulses. %An example of such shaped pulse is displayed in Fig.~\ref{fig:comparison} for a linearly ordinary polarized input visible pulse with a pulse duration of 15 ps. The group velocities mismatch leads to a periodic modulation of the polarization state of the visible pulse by the THz pulse at THz frequencies as shown by Fig.~\ref{fig:comparison} a)-d) for different times t.  
As the modulation frequency can be modified easily it can be tuned to resonantly drive systems like coherent THz optical phonon excitation in solids~\cite{sokolowski2003femtosecond,fritz2007ultrafast} and Raman-active media~\cite{misawa2016applications} for example. Another possible application of our experimental scheme is to add the third (axial) coordinate to the well-known "2D imaging with THz waves"~\cite{hu1995imaging,wu1996two,nuss1996chemistry,fitzgerald2002introduction}, which  currently provides only lateral resolution. The micrometer axial resolution for depth-profiling of the suggested three-dimensional imaging technique is controlled by the spatial overlap of the THz and probe light pulses and, in case of their counter-propagation, could outperform the "echography with THz pulses"~\cite{jackson2015terahertz}. 

In conclusion, we have experimentally demonstrated an
ultrafast tunable polarization modulation of optical pulses at a
wavelength of 400 and 800 nm caused by nonresonant interaction
with an ultrashort THz pulse in LiNbO$_3$ crystal. The
maximum polarization rotation and ellipticity was measured
to be $\sim 6\%$ and $\sim 8\%$ at the internal angle of $\sim 8.5^\circ$ between
the THz pulse and a 400 nm visible probe. This effect is modelled
by solving the wave propagation equation, and analytical
expressions show that an interplay between distinct phase- and
group-matching conditions for visible and THz electromagnetic
fields determines the overall conversion efficiency. Being
in excellent agreement with theoretical calculations, our experimental
findings pave the way to the design of ultrafast and
efficient miniaturized electro-optical devices for pulse shaping
operating in the multi-THz range.

\section*{Funding}
Campus France; Région Pays de la Loire (NNN-Telecom, Ultrasfast Acoustics in Hybrid Magnetic Nanostructures and Echopico projects).

% Bibliography

% Full bibliography added automatically for Optics Letters submissions; the following line will simply be ignored if submitting to other journals.
% Note that this extra page will not count against page length

\bibliographystyle{ieeetr}
\bibliography{sample}

\end{document}